\documentclass[12pt,showpacs]{revtex4}
\usepackage{graphicx,amsmath,amssymb}

\newcommand{\beq}{\begin{equation}}
\newcommand{\eeq}{\end{equation}}
\newcommand{\beqa}{\begin{eqnarray}}
\newcommand{\eeqa}{\end{eqnarray}}
\newcommand{\ket}[1]{| #1 \rangle}
\newcommand{\bra}[1]{\langle #1 |}

\begin{document}


\title{Relative quantum phase, $m$-tangle, and  multi-local Lorentz-group invariant}

\author{Hoshang Heydari}
\email{hoshang@physto.se} \affiliation{Physics Department, Stockholm university, 10691 Stockholm Sweden}

\date{\today}

\begin{abstract}
In this paper we establish a relation between quantum relative  phase, $m$-tangle, and  multi-local Lorentz-group invariant or $SL(2,\mathbb{C})^{\times m}$-invariant $S^{2}_{(m)}$. Our construction is based on the
orthogonal complement of a  positive
operator valued measure on quantum phase. In particular, we propose a quantity  based on the quantum relative phase of a multi-qubit operator that coincides with
 $m$-tangle, and  multi-local Lorentz-group invariant.
\end{abstract}
\maketitle
\section{Introduction}
Quantum entanglement is an interesting quantum phenomena, with many applications in the field of quantum information processing. The problem of quantifying and classifying multipartite quantum systems is a challenging task which e.g., could results in designing many powerful quantum algorithms. Recently, there have been an increase of activity among researcher to  construct measures of entanglement for bipartite and  multipartite systems.  One of the well-known measures of entanglement for
 a pair of qubits is  the concurrence, which is directly related to
the entanglement of formation
\cite{Bennett96,Wootters98,Wootters00}. For multi-qubit system, we have also some important measures of entanglement such as $m$-tangle,  multi-local Lorentz-group invariant, and Hilbert-Schmidt distance \cite{Wong01,Jaeger03,Jaeger07}.
 Recently, we have also defined concurrence
classes for  multi-qubit  states \cite{Hosh5} based on an
orthogonal complement of a  positive
operator valued measure (POVM) on quantum phase. In this paper, we will also construct a quantity which coincides with $m$-tangle and multi-local Lorentz-group invariant  based on orthogonal complement of a POVM. In particular, in section  \ref{mtangle} we review the construction of $m$-tangle, multi-local Lorentz-group invariant, and Hilbert-Schmidt distance. We also discuss in detail the relation between these measures of entanglement. In section \ref{con1} we will first review the construction of the POVM. Then, we will  construct a measure of entanglement by considering a POVM  that includes all subsystems quantum phases of a multi-partite systems. Finally, we will establish a relation between this measure of entanglement and
$m$-tangle, multi-local Lorentz-group invariant, and Hilbert-Schmidt distance.


\section{M-tangle, multi-local Lorentz-group invariant, and  Hilbert-Schmidt distance}\label{mtangle}
 Here we will give a short introduction to the concurrence and one of its generalization, namely the $m$-tangle. We will also review the construction of  multi-local Lorentz-group invariant and Hilbert-Schmidt distance. Moreover, we will discuss the relation between these measures of entanglement.
Now, we define a  multi-qubit state as
\begin{equation}\ket{\Psi}=\sum^{1}_{x_{m-1},x_{m-2},\ldots,
x_{0}=0}\alpha_{x_{m-1}x_{m-2}\cdots x_{0}}\ket{x_{m-1}x_{m-2}\cdots
x_{0}},
\end{equation}
 with corresponding  Hilbert space $
\mathcal{H}_{\mathcal{Q}}=\mathcal{H}_{\mathcal{Q}_{1}}\otimes
\mathcal{H}_{\mathcal{Q}_{2}}\otimes\cdots\otimes\mathcal{H}_{\mathcal{Q}_{m}}
$.  For example, a pure
two-qubit state is give by $\ket{\Psi}=\sum^{1}_{x_{1},
x_{0}=0}\alpha_{x_{1}x_{0}}\ket{x_{1}x_{0}}\in \mathcal{H}_{\mathcal{Q}_{1}}\otimes
\mathcal{H}_{\mathcal{Q}_{2}}=\mathbb{C}^{2}\otimes\mathbb{C}^{2}$.
Moreover, let us
introduce a complex conjugation operator $\mathcal{C}_{m}$ that
acts on the multipartite quantum state $\ket{\Psi}$  as
\begin{equation}\label{cong}
\ket{\Psi^{*}}=\mathcal{C}_{m}\ket{\Psi}=\sum^{1}_{x_{m-1},x_{m-2},\ldots,
x_{0}=0}\alpha^{*}_{x_{m-1}x_{m-2}\cdots x_{0}}\ket{x_{m-1}x_{m-2}\cdots
x_{0}}
.
\end{equation}
 The
density operator $\rho_{\mathcal{Q}}$ is said to be fully
separable, which we will denote by $\rho^{sep}_{\mathcal{Q}}$,
with respect to the Hilbert space decomposition, if it can  be
written as $ \rho^{sep}_{\mathcal{Q}}=\sum^\mathrm{N}_{n=1}p_{n}
\bigotimes^m_{j=1}\rho^{n}_{\mathcal{Q}_{j}}$,
$\sum^\mathrm{N}_{n=1}p_{n}=1 $,
 for some positive integer $\mathrm{N}$, where $p_{n}$ are positive real
numbers and $\rho^{n}_{\mathcal{Q}_{j}}$ denote a density operator
on Hilbert space $\mathcal{H}_{\mathcal{Q}_{j}}$. If
$\rho^{p}_{\mathcal{Q}}$ represents a pure state, then the quantum
system is fully separable if $\rho^{p}_{\mathcal{Q}}$ can be
written as
$\rho^{sep}_{\mathcal{Q}}=\bigotimes^m_{j=1}\rho_{\mathcal{Q}_{j}}$,
where $\rho_{\mathcal{Q}_{j}}$ is a density operator on
$\mathcal{H}_{\mathcal{Q}_{j}}$. If a state is not separable, then
it is called an entangled state.

The concurrence of two-qubit states is defined as
 \begin{equation}\mathcal{C}(\Psi)=|\langle\Psi\ket{\widetilde{\Psi}}|^{2},
 \end{equation}
  where  the tilde represents the "spin-flip" operation
 $\ket{\widetilde{\Psi}}=\sigma_{y}\otimes
  \sigma_{y}\ket{\Psi^{*}}$
 and $\ket{\Psi^{*}}$ is defined by equation (\ref{cong}) and $\sigma_{y}=\left(%
\begin{array}{cc}
  0 & -i \\
  i & 0 \\
\end{array}%
\right)$ is a Pauli spin-flip operator
\cite{Wootters98,Wootters00}. This construction can be generalized to a multi-qubit system by defining
\begin{equation}\ket{\widetilde{\Psi}}=\sigma^{\otimes m}_{y}
\ket{\Psi^{*}},
\end{equation}
where $\sigma^{\otimes m}_{y}$ denotes $m$-folds tensor product of $\sigma_{y}$. Next, we define $m$-tangle as
 \begin{equation}\tau_{m}=|\langle\Psi\ket{\widetilde{\Psi}}|^{2}
 \end{equation}
for every even $m$-qubit system \cite{Wong01}. The $m$-tangle  is a symmetry based measure of entanglement. Next, we define a multi-local Lorentz-group invariant or $SL(2,\mathbb{C})^{\times m}$-invariant $S^{2}_{(m)}$  by \cite{Jaeger03}
 \begin{equation}S^{2}_{(m)}(\rho)=\mathrm{Tr}(\rho\widetilde{\rho}),
 \end{equation}
where the generalized spin-flip operation is defined by
\begin{equation}
\widetilde{\rho}=\sigma^{\otimes m}_{y}\rho^{*}\sigma^{\otimes m}_{y}.
 \end{equation}
The multi-local Lorentz-group invariant is also related to the $m$-tangle as follows
 \begin{equation}
 S^{2}_{(m)}(\ket{\Psi})=\tau_{m}=|\langle\Psi\ket{\widetilde{\Psi}}|^{2}.
 \end{equation}
 Moreover, there is a relation between Hilbert-Schmidt distance and the multi-local Lorentz-group invariant as follows
  \begin{equation}
  S^{2}_{(m)}(\ket{\Psi})=P(\rho)-D^{2}_{HS}(\rho-\widetilde{\rho}),
 \end{equation}
 where $P(\rho)$ is the purity of $\rho$ and $D^{2}_{HS}(\rho-\widetilde{\rho})
 =\frac{1}{\sqrt{2}}\left(\mathrm{Tr}(\rho-\widetilde{\rho})^{2}\right)^{1/2}$. Furthermore, there is a relation between the spin-flip symmetry measure which is defined as
  \begin{equation}
  I(\rho,\widetilde{\rho})=1-D^{2}_{HS}(\rho-\widetilde{\rho})=1+S^{2}_{(m)}(\rho)-P(\rho)
 \end{equation}
 and the multi-local Lorentz-group invariant and  between Hilbert-Schmidt which are also related to the $m$-tangle for pure state.
In the following section we will show that these quantities are related to quantum phase of a multi-qubit state.

\section{Relative quantum phase and  $m$-tangle}\label{con1}
 In this section we will establish a relation between the multi-local Lorentz-group invariant, Hilbert-Schmidt, $m$-tangle,  and a measure of entanglement for multi-qubit states based on the orthogonal complement of the POVM on quantum phase.
Our POVM is a set of linear operators
$\Delta(\varphi_{1,2},\ldots,\varphi_{1,N},\varphi_{2,3},\ldots,\varphi_{N-1,N})$
furnishing the probabilities that the measurement of a state
$\rho$ on the Hilbert space $\mathcal{H}$ is given by
\begin{equation}
\mathrm{p}(\varphi_{1,2},\ldots,\varphi_{1,N},\varphi_{2,3},\ldots,\varphi_{N-1,N})=
\mathrm{Tr}(\rho\Delta(\varphi_{1,2},\ldots,\varphi_{1,N},\varphi_{2,3},\ldots,\varphi_{N-1,N})),
\end{equation}
where
$(\varphi_{1,2},\ldots,\varphi_{1,N},\varphi_{2,3},\ldots,\varphi_{N-1,N})$
are the outcomes of the measurement of the quantum  phase, which is
discrete and binary. This POVM satisfies the following properties,
$\Delta(\varphi_{1,2},\ldots,\varphi_{1,N},\varphi_{2,3},\ldots,\varphi_{N-1,N})$
is self-adjoint, is positive, and  is normalized, i.e.,
\begin{equation}\overbrace{\int_{2\pi}\cdots
\int_{2\pi}}^{N(N-1)/2}d\varphi_{1,2}\cdots
d\varphi_{1,N}d\varphi_{2,3}\cdots d\varphi_{N-1,N}
    \Delta(\varphi_{1,2},\ldots,\varphi_{N-1,N})=\mathcal{I},
\end{equation}
where the integral extends over any $2\pi$ intervals of the form
$(\varphi_{k},\varphi_{k}+2\pi)$ and $\varphi_{k}$ are the
reference phases for all $k=1,2,\ldots,N$.
A general and symmetric POVM in a single $N_{j}$-dimensional
Hilbert space $\mathcal{H}_{\mathcal{Q}_{j}}$ is given by
\begin{equation}
\Delta(\varphi_{k_{j},l_{j}})=
\sum^{N_{j}}_{l_{j}=1}\sum^{N_{j}}_{k_{j}=1}
e^{i\varphi_{k_{j},l_{j}}}\ket{k_{j}}\bra{l_{j}},
\end{equation}
where $j=1,2,\ldots,m$, $\ket{k_{j}}$ and $\ket{l_{j}}$ are the basis vectors in
$\mathcal{H}_{\mathcal{Q}_j}$ and the quantum phases satisfy the
following relation $ \varphi_{k_{j},l_{j}}=
-\varphi_{l_{j},k_{j}}(1-\delta_{k_{j} l_{j}})$. Moreover, the
orthogonal complement of our POVM
 is given by
\begin{equation}
\widetilde{\Delta}_{\mathcal{Q}_{j}}(\varphi_{k_{j},l_{j}})=\mathcal{I}_{N_{j}}-
\Delta_{\mathcal{Q}_{j}}(\varphi_{k_{j},l_{j}}),
\end{equation}
 where
$\mathcal{I}_{N_{j}}$ is the $N_{j}$-by-$N_{j}$ identity matrix for
subsystem $j$
\cite{Hosh5}. The POVM is a function of the $N_{j}(N_{j}-1)/2$
phases
$(\varphi_{1_j,2_j},\ldots,\varphi_{1_j,N_j},\varphi_{2_j,3_j},\ldots,\varphi_{N_{j}-1,N_{j}})$.
It is now possible to form a POVM of a multipartite system by
simply forming the tensor product
\begin{eqnarray}\label{POVM}\nonumber
\Delta_\mathcal{Q}(\varphi_{k_{1},l_{1}},\ldots,
\varphi_{k_{m},l_{m}})&=&
\Delta_{\mathcal{Q}_{1}}(\varphi_{k_{1},l_{1}})\otimes\cdots
\otimes
\Delta_{\mathcal{Q}_{m}}(\varphi_{k_{m},l_{m}}),
\end{eqnarray}
where, e.g., $\varphi_{k_{1},l_{1}}$ is the set of
POVMs phase associated with subsystems $\mathcal{Q}_{1}$, for all
$k_{1},l_{1}=1,2,\ldots,N_{1}$, where we need only to consider
when $l_{1}>k_{1}$.
 The
unique structure of our POVM enables us to distinguish different
classes of multipartite states. In the $m$-partite case, the  off-diagonal elements of
the matrix corresponding to
\begin{equation}
\widetilde{\Delta}_\mathcal{Q}(\varphi_{k_{1},l_{1}},\ldots,
\varphi_{k_{m},l_{m}})=
\widetilde{\Delta}_{\mathcal{Q}_{1}}(\varphi_{k_{1},l_{1}})
\otimes\cdots
\otimes\widetilde{\Delta}_{\mathcal{Q}_{m}}(\varphi_{k_{m},l_{m}}),
\end{equation}
 have phases that are sum
or differences of phases originating from two and $m$ subsystems.
That is, in the later case the phases of
$\widetilde{\Delta}_\mathcal{Q}(\varphi_{k_{1},l_{1}},\ldots,
\varphi_{k_{m},l_{m}})$ take the form
$(\varphi_{k_{1},l_{1}}\pm\varphi_{k_{2},l_{2}}
\pm\ldots\pm\varphi_{k_{m},l_{m}})$.
 For example for the case when the phases originating from  $m$ subsystems, we define a linear
operators based on our POVM which are sum and difference of phases
of $m$-subsystems by
\begin{eqnarray}\nonumber
\widetilde{\Delta}_{(m)}(\varphi_{k_{1},l_{1}},\ldots,
\varphi_{k_{m},l_{m}})
&=&\widetilde{\Delta}_{\mathcal{Q}_{1}}
(\varphi_{k_{1},l_{1}})
\otimes\widetilde{\Delta}_{\mathcal{Q}_{2}}
(\varphi_{k_{2},l_{2}})
\otimes\cdots\otimes\widetilde{\Delta}_{\mathcal{Q}_{2}}
(\varphi_{k_{m},l_{m}})\\\nonumber&\equiv&\mathrm{antidiag}(e^{i(\varphi_{k_{1},l_{1}}
+\varphi_{k_{2},l_{2}}+\cdots+\varphi_{k_{m-1},l_{m-1}}
+\varphi_{k_{m},l_{m}})},\\\nonumber&&e^{i(\varphi_{k_{1},l_{1}}
+\varphi_{k_{2},l_{2}}+\cdots+\varphi_{k_{m-1},l_{m-1}}
-\varphi_{k_{m},l_{m}})},\ldots,\\\nonumber&&e^{i(-\varphi_{k_{1},l_{1}}
-\varphi_{k_{2},l_{2}}-\cdots-\varphi_{k_{m-1},l_{m-1}}
+\varphi_{k_{m},l_{m}})},\\&&e^{i(-\varphi_{k_{1},l_{1}}
-\varphi_{k_{2},l_{2}}-\cdots-\varphi_{k_{m-1},l_{m-1}}
-\varphi_{k_{m},l_{m}})})
\end{eqnarray}
where by choosing
$\varphi_{k_{j},l_{j}}=\frac{\pi}{2}$ for all
$k_{j}<l_{j}, ~j=1,2,\ldots,m$, we get an operator which has the
structure of Pauli operator $\sigma_{y}$ embedded in a
higher-dimensional Hilbert space and coincides with $\sigma_{y}$
for a single-qubit. Next, we will define following quantity
 \begin{equation}\Gamma_{m}(\ket{\Psi})=|\langle\Psi\ket{\widetilde{\Delta}(\varphi_{k_{1},l_{1}},\ldots,
\varphi_{k_{m},l_{m}})\Psi^{*}}|^{2},
 \end{equation}
where $\ket{\Psi^{*}}$ is given by equation  (\ref{cong}). For multi-qubit state, that is $k_{j}<l_{j}=2, ~j=1,2,\ldots,m$,  the operator $\widetilde{\Delta}(\varphi_{k_{1},l_{1}},\ldots,
\varphi_{k_{m},l_{m}})$ reduced to
\begin{eqnarray}
\widetilde{\Delta}_{(m)}(\varphi_{k_{1},l_{1}},\ldots,
\varphi_{k_{m},l_{m}})
&=&\sigma_{y}
\otimes\sigma_{y}
\otimes\cdots\otimes\sigma_{y}=\sigma^{\otimes m}_{y}.
\end{eqnarray}
Thus, we have established the following relation between $\Gamma_{m}(\ket{\Psi})$, the $m$-tangle, Hilbert-Schmidt distance and the multi-local Lorentz-group invariant
\begin{equation}
\Gamma_{m}(\ket{\Psi})=\tau_{m}= S^{2}_{(m)}(\ket{\Psi})=P(\rho)-D^{2}_{HS}(\rho-\widetilde{\rho})
\end{equation}
for even $m$ and $m>3$ by definition.
For mixed multi-qubit system we have
 \begin{equation}\Gamma_{m}(\rho)
 =\mathrm{Tr}(\rho\widetilde{\Delta}_{(m)}(\varphi_{k_{1},l_{1}},\ldots,
\varphi_{k_{m},l_{m}})\rho^{*}\widetilde{\Delta}_{(m)}(\varphi_{k_{1},l_{1}},\ldots,
\varphi_{k_{m},l_{m}}))=S^{2}_{(m)}(\rho)
 \end{equation}
is  written in terms of the generalization of the spin-flip. As an example we consider the GHZ state $\ket{\Psi_{GHZ}}=\frac{1}{\sqrt{2}}(\ket{0}^{\otimes m}+\ket{1}^{\otimes m})$. For this state we have $\Gamma_{m}(\ket{\Psi_{GHZ}})=2|-\frac{1}{4}-\frac{1}{4}|=\tau_{m}=1$.

These results illustrate the importance of the quantum relative phase of multi-partite systems in very concrete way. The advantages of the quantity $\Gamma_{m}$ are the following. First of all it has a very good physical interpretation to be a measure of entanglement in terms of relative phase of multi-partite quantum system and for the second it can also be generalized to a general multi-partite system.

\begin{flushleft}
\textbf{Acknowledgments:} The  work was supported  by the Swedish Research council (VR).
\end{flushleft}



\begin{thebibliography}{99}
\bibitem{Bennett96} C. H. Bennett, D. P. DiVincenzo, J. Smolin, and W.~K.~Wootters,
Phys. Rev. A {\bf 54}, 3824 (1996).
\bibitem{Wootters98} W. K. Wootters,  Phys. Rev. Lett. {\bf 80}, 2245 (1998).
\bibitem{Wootters00} W. K. Wootters, Quantum Information and Computaion, Vol. 1, No. 1 (2000) 27-44, Rinton Press.
\bibitem{Wong01} A. Wong and N. Christensen Phys. Rev. A {\bf 63}, 044301 (2001).
\bibitem{Jaeger03} G. Jaeger, V. Sergienko, A. Saleh, and C. Teich, Phys. Rev. A {\bf 68}, 022318 (2003).
\bibitem{Jaeger07} G. Jaeger, {\it Quantum Information  An Overview }, Springer, New York 2007.
\bibitem{Hosh5} H. Heydari, J. Phys. A: Math. Gen. 38 (2005) 11007-11017.





\end{thebibliography}
\end{document}